\documentclass{elsart}

\usepackage{graphicx}% Include figure files
\usepackage{dcolumn}% Align table columns on decimal point
\usepackage{epsf}

\newcommand{\beq}{\begin{equation}}
\newcommand{\eeq}{\end{equation}}
\newcommand{\bea}{\begin{eqnarray}}
\newcommand{\eea}{\end{eqnarray}}

\begin{document}
\begin{frontmatter}

%\vskip 1.0truecm

\begin{flushright}
{UCLA/05/TEP/23}
\end{flushright}

\title{Interactions of ultrahigh-energy cosmic rays with photons in the
galactic center} 

\author{Alexander Kusenko$^a$, Jason Schissel$^a$, and Floyd
W. Stecker$^{a,b}$} 

%\affiliation{
\address
%\thanks
{$^a$ Department of Physics and Astronomy,
UCLA, Los Angeles, CA 90095-1547, USA \\ $^b$Laboratory for High Energy
Astrophysics, NASA Goddard Space Flight Center, Greenbelt, MD 20771 } 
%\vspace{0.5truecm}

\begin{abstract}
Ultrahigh-energy cosmic rays passing through the central region
of the Galaxy interact with starlight and the infrared photons.  Both nuclei
and protons generate secondary fluxes of photons and neutrinos on their passage
through the central region. We compute the fluxes of these secondary particles,
the observations of which can be used to improve one's understanding of
origin and composition of ultrahigh-energy comic rays, especially if the
violation of the Greisen--Zatespin--Kuzmin cutoff is confirmed by the
future data. 
 
\end{abstract}

\end{frontmatter}

Ultrahigh-energy cosmic rays (UHECR) can interact with the cosmic microwave
background radiation (CMBR) photons and produce pions.  This process, which is
the main source of energy losses for the highest-energy cosmic rays, is
supposed to result in the Greisen--Zatsepin--Kuzmin (GZK) cutoff~\cite{gzks}.
However, observations of ultrahigh-energy cosmic rays (UHECR) show a number of
events with energies above $10^{20}$eV~\cite{data}.  While the first data
reported by Pierre Auger experiment~\cite{Sommers:2005vs} neither confirm, nor
rule out the violation of the GZK cutoff reported by AGASA, one expect much
more definitive results in the near future. If UHECR interact with the CMBR,
then GZK cutoff will soon be observed and the photons from pion decays should
also be discovered in the near future~\cite{gks05}. 

If, however, the cosmic ray spectrum continues beyond $10^{20}$eV without GZK
suppression, then either the flux of UHECR is dominated by nearby sources (for
example, decaying superheavy relic particles~\cite{particles,kt}), or the
photomeson interactions with the CMB photons are stymied by some new physics,
for example, a violation of the Lorentz invariance~\cite{lorentz}.  In either
case, the diffuse flux of UHE photons is either small or zero.  However, in
either case, the UHECR protons and nuclei can interact with the photons in the
galactic center (GC), where the density of photons is very high, and where the
average energy of photons is much higher than that of the CMB photons. 
Detection of photons or neutrinos from UHECR interactions in the galactic
center, in the absence of GZK cutoff and diffuse UHE photon flux would be an
important indication of new physics.  

In this paper we examine the propagation of UHECR through the central region
of Galaxy which contains a relatively high density of starlight photons,
infrared (IR) photons, and interstellar gas.  

% There are several reasons
% why it is desirable to understand the effect of galactic propagation of UHECR
% from remote sources though the galactic center region. Some models of 
% UHECR predict a higher flux from the galactic center.  For example, if UHECR 
% originate from decaying superheavy dark matter particles, one expects an 
% anisotropy in the arrival directions of UHECR and particularly UHE photons, 
% with the arrival direction distribution peaking at the galactic center (GC). 
% One also expects an additional flux of UHE photons produced by the decays
% of pions from interactions of 
% UHECR protons and nuclei with the starlight and IR photons that have an
% enhanced density in the GC. Fluxes and spectra from such photomeson
% interactions depend 
% on whether the UHECR primaries are protons or nuclei.
% So, in principle, the UHE photon spectra from the GC can be used to determine
% the nuclear composition of UHECR coming from direction of the GC.

The energy and composition of extragalactic cosmic rays passing through the 
central region of Galaxy can be altered by their interactions with 
starlight photons and infrared photons emitted by dust reradiation of
starlight. Interactions of ultrahigh protons with such photons results 
in the production of pions~\cite{gzks} which generate a secondary 
flux of photons and neutrinos~\cite{st79}. 
The most important interactions involving ultrahigh energy nuclei are
photodisintegration interactions~\cite{st69}, similar to the 
Zatsepin--Gerasimova effect for interactions with solar photons
\cite{zg,Medina-Tanco:1998ac,Epele:1998mv}.

These interactions can be observed in different ways. First, there will be a
suppression in UHECR observed in the direction of the central galaxy, but this
shadow may be difficult to observe and identify. The detection of
UHE photons and neutrinos from interactions of nuclei and protons in the
region of the GC presents a more promising study of UHECR. Also,
photodisintegration of nuclei can decrease the average atomic weight of UHECR
nuclei coming from the direction of the GC, so that one can look for such
a change in composition.

Let us now discuss the interactions of UHECR nuclei and protons with 
starlight.  One can model the photon density in the galactic
center using a stellar population model based on star counts~\cite{bahcall}.
In reality the distribution of stars is more complicated, non-uniform, with
bright clusters~\cite{clusters} and gaps between them.  However, since these
bright clusters do not present extensive optically thick targets, we can
use a smoothed-out stellar distribution.  We assume that all stars have the
same average luminosity $L_*$.  Since the angular size of the central core
region is not much bigger than the angular resolution of UHECR experiments, 
we can consider the total photon distribution to be approximately
spherical.  Let us denote the number density of stars as $n_*(r)$ and the
number density of photons as $n_\gamma(r)$.  The total number of photons
passing through a sphere of radius $R$ centered at the GC per unit time is
equal to the total number of photons produced inside such a sphere,
$I_1(R)$, plus the photons originating outside the sphere
and passing through it, $I_2(R)$: 

\bea
I_1 & =& \int_0^R 4 \pi r^2 L_* n_*(r) dr\\
I_2 & = & \int_{R}^\infty  4 \pi r^2 L_* n_*(r) \left( 1-\sqrt{1-R^2/r^2}
\right) dr
\eea
 The same number of photons can be written as $4\pi R^2 \times n_\gamma
(R)$.   From the equality of these two fluxes we get an estimate for the
number density of photons produced by a given distribution of stars:
\begin{equation}
n_\gamma (R) = \frac{L_*}{R^2} \int_0^\infty n_*(r) f(r) r^2 dr, 
\end{equation}
where
\begin{equation}
f(r) = \left\lbrace \begin{array}{ll}
1, & r<R \\
1-\sqrt{1-R^2/r^2}, & r\ge R
\end{array} \right .
\label{ngamma}
\end{equation}
One can approximate the stellar density as  
\beq
n_*(r)=n_0 r^{-1.8} \exp \{
(-r/{\rm kpc})^3 \}, 
\label{nstar}
\eeq
where $n_0\approx 0.8 \times 10^6 {\rm pc}^{-3}$~\cite{bahcall}.   
This estimate could be further improved if
one needed to take into account the angular distribution at angles much smaller
than a degree.  One could, for example, use astronomical data from MSX and IRAS
surveys and try to reconstruct the photon density based on the photometry data
of specific regions in the vicinity of the GC.  However, for our purposes the
estimate given by eq. \ref{ngamma} is sufficient because we are interested in
the effect on
UHECR spectrum and composition integrated over approximately one square degree 
around GC.

The density of IR photons in the Central Molecular Zone
(CMZ)~\cite{morris} is the highest in dense clouds of dust, in which the
starlight is absorbed and remitted as the infrared light.  Gas and dust in
the CMZ have temperatures ranging from 30~K to 200~K, with an average
temperature of 70~K~\cite{morris,IR}.  We make a simplified model of the IR
radiation field near the galactic center by assuming that all the IR photons
come from a spherical dust cloud with radius $\approx 5$pc, centered near GC. 
The spectrum of IR photons is assumed to be thermal, with temperature 70~K. 

% 
% The gas clouds present in the CMZ provide 
% a target for $pp$-interactions as well, with a much larger cross
% section than $p\gamma$ interactions, but the density of the CMZ is too low to
% give a significant contribution to the photon flux.

Other sources of photons are present in the central region of the Galaxy, 
but they do not give a significant contribution. 
For example, a few tens of supernovae happen during the passage of a cosmic
ray through the galactic center region, but only those cosmic rays that pass
closer
than a light-year away from a supernova within the first year since its
explosion
can interact efficiently with the supernova photons.  We estimate that this
has a negligible effect on the overall flux.

\begin{figure}
\centerline{\epsfxsize=5in\epsfbox{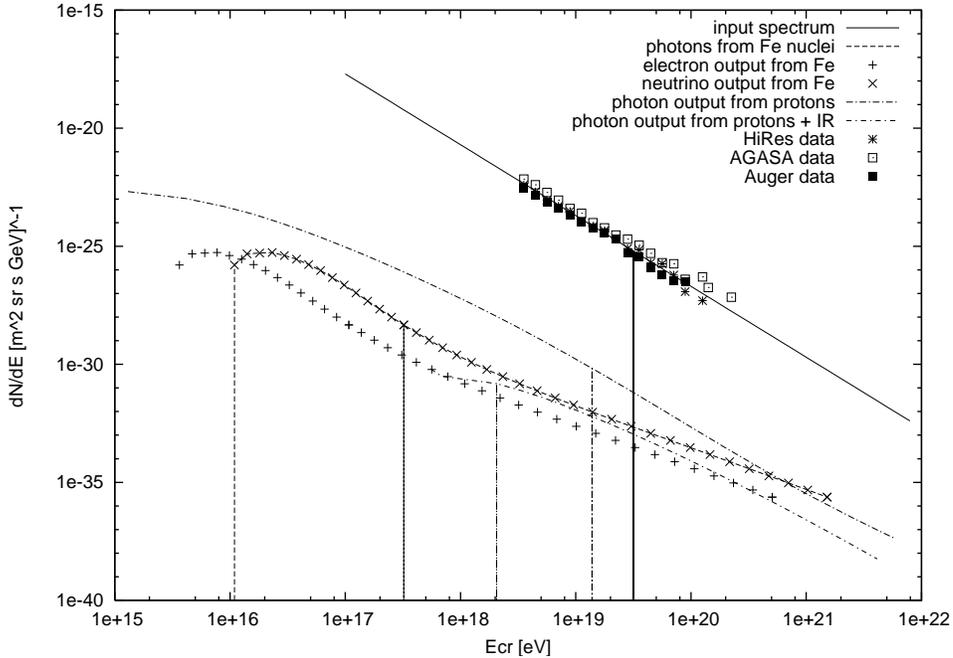}}   
%\centerline{\epsfxsize=3.5in\epsfbox{iron_E.ps}}   
\caption{ The spectrum of secondary photons produced by cosmic ray
interactions near the galactic center (GC).  The fluxes are averaged over a
square degree near the GC. The dashed line represents the input spectrum of
UHECR, assumed to be a simple power-law without a GZK cutoff ({\em cf.}
Fig.~2).  Also shown are the data points from AGASA, HiRes, and Pierre Auger
experiments. (These points are the reorted central values drawn to guide the
eye; the error bars are not shown.)  The solid and the dotted line show the
spectra of secondary photons assuming the primaries are protons or iron nuclei,
respectively.   As discussed in the text, the predicted spectrum of high-energy
neutrinos is very close to that of photons. 
}
\label{fig_spectrum}
\end{figure}

\begin{figure}
\centerline{\epsfxsize=5in\epsfbox{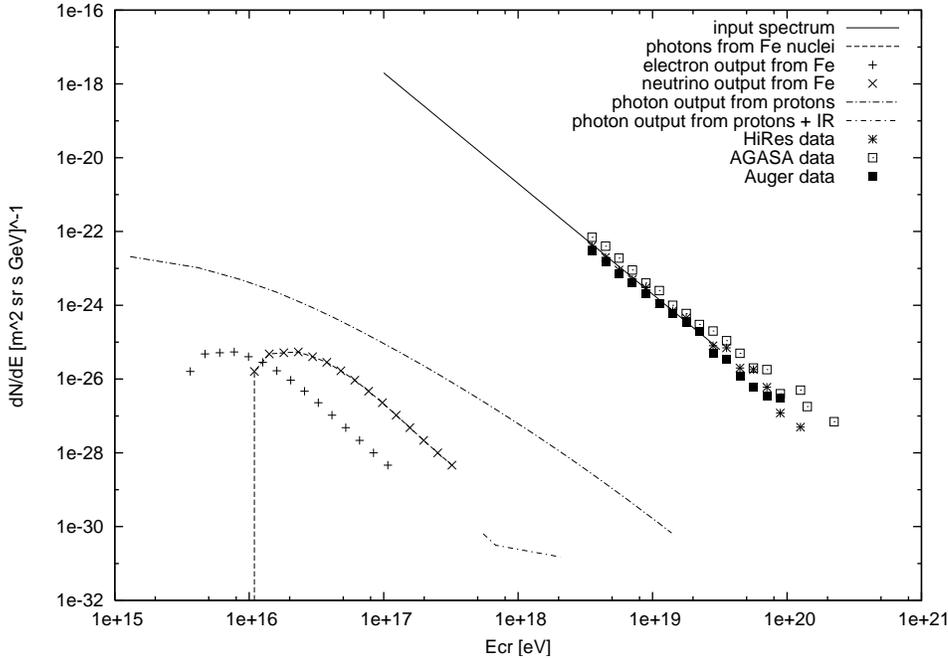}}   
%\centerline{\epsfxsize=3.5in\epsfbox{iron_E.ps}}   
\caption{ The spectrum of secondary photons in the case of a sharp GZK cutoff
for the input spectrum.  If photomeson interactions do take place and produce 
GZK cutoff, the diffuse photon background exceeds the flux shown here.  As in
Fig.~1, the fluxes are averaged over a square degree near the GC. The dashed
line represents the input spectrum of UHECR, assumed to be a simple power-law. 
The solid and the dotted line show the spectra of secondary photons assuming
the primaries are protons or iron nuclei, respectively.   
}
\label{fig_spectrum_cutoff}
\end{figure}

We have computed numerically the spectra of photons produced by the cosmic rays
passing through the central bulge under the assumption that the primaries are
(i) protons and (ii) iron nuclei.  In reality, one should probably expect the
composition to be a mixture of different nuclei, unless the sources are such
that they cannot produce UHE nuclei at all (this is the case in top-down
scenarios, for example).  The resulting spectra are shown in
Fig.~\ref{fig_spectrum}.  The injection spectrum of UHECR is assumed to be a
simple power-law spectrum, consistent with AGASA and Pierre Auger
results~\cite{data,Sommers:2005vs}. Since the photons from the galactic center
are of most interest if the GZK cutoff is \textit{not} detected, we have
assumed no suppression of UHECR flux at energies beyond $10^{20}$~eV for
fluxes shown in Fig.~\ref{fig_spectrum}.  In Fig.~\ref{fig_spectrum_cutoff} we
show our results for the input spectrum that has a GZK cutoff.

The photon field near GC is sufficiently thin and is sufficiently close to
Earth that one need not include a full cascade calculation.  If an UHECR proton
interacts with a starlight photon, it produces a pion and either a proton or a
neutron in the final state.   

The flux of secondary protons is too low to be of
interest.  Secondary neutrons at these energies do not have enough time to
decay.  They arrive at Earth unimpeded by cosmic background and undeflected by
magnetic fields.  If some experiments could distinguish between protons and
neutrons, the galactic center would be seen as a source of neutrons.  
However, both
techniques used for the detection of UHECR, the ground array and the
fluorescent telescopes, are unable to distinguish a shower started by a
neutron from the one started by a proton. Therefore, only the photons and the
neutrinos are of interest to us, and only a single interaction of UHECR hadron
with a photon needs to be considered.

The spectrum of neutrinos is very close to that of photons shown in
Fig.~\ref{fig_spectrum}. Indeed, one-third of pions produced in photomeson
interactions are $\pi^0$, and they produce two photons each when they decay.
The other two-third of the pions produced in these reactions generate one muon
and one muon neutrino each.  These neutrinos have energy spectrum similar to
that of photons from $\pi^0$ decays.  The muon decays produce additional
neutrinos at lower energies, but at lower energies they energies they give a
small contribution to the neutrino flux. As a result, the high-energy neutrinos
have a spectrum that is very close to that of the photons in
Fig.~\ref{fig_spectrum}.  Both the uncertainties in the input spectrum and the
experimental uncertainties are much larger than the difference between the two
fluxes.  

In addition to photons and neutrinos, pion decays produce electrons.
Experimentally one probably cannot distinguish between atmospheric showers
initiated by photons and electrons.  As shown in Fig.~\ref{fig_spectrum}, the
electrons do not give an appreciable contribution to the photon flux.

% 
% {\bf *** check the status of this }
% {\em 
% There are also three clusters observed recently that would be large targets
% for UHECR.
% Hubble Space Telescope/NICMOS Observations of Massive Stellar Clusters Near
% the Galactic Center, D.F. Figer, S.S. Kim, M. Morris, E. Serabyn, R.M. Rich,
% and I.S. McLean, Ap.J., 525, 750  758 (1999).
%   }
%   {\bf ***} 

The optical depth of the galactic center region is less than one.  If the
photomeson interactions of UHECR do take place, then the optical depth of the
universe is much greater than one, and the isotropic extragalactic flux of UHE
photons~\cite{gks05} exceeds the contribution from the galactic center. 
However, in the absence of GZK cutoff caused by photomeson interactions,  the
photons from the galactic center dominate.  The angular size of the photomeson
region is about  $\Omega=0.03 $~sr.  Hence, one expects several events per year
in Pierre Auger.  

For comparison, we also show, in Fig.~\ref{fig_spectrum_cutoff} the fluxes of
photons and neutrinos in the case when the input specturm is suppressed for 
energies beyond the GZK cutoff.  Of course, in this case one expects a
stronger diffuse photon flux. 

% We can compare the photon fluxes in Figure 1 to isotropic extragalactic 
% background photons integrated over the angular size of the GC region.
% The photons expected from photomeson interactions of UHECRs with the
% 2.7 K cosmic background radiation will be derived from the calculations
% given in Ref.~\cite{gks05}. There are also background fluxes predicted
% from various ``top-down'' models *REFERENCES* which we normalize from
% the upper limits obtained by AGASA as in Ref.~\cite{gks05}.

% 
% MORE DISCUSSION AND ANOTHER FIGURE
% 
% Thus we expect Auger to be able to observe a GC flux ? sigma above
% the photomeson photon background and ? sigma above a possible ``top-down''
% background. The expected GC photon event rate for Auger is ?????.
% 
% GC NEUTRINOS - I GIVE THE NEUTRINO SENSITIVITY FOR OWL IN MY CATANIA PAPER 
% 
% OBSERVED COMPOSITION CHANGES? AUGER? OWL?

One can envision several ways in which the observations of GC can be used to
understand the origin and composition of UHECR.  If the violation of the GZK
cutoff is confirmed by Pierre Auger experiment, one can look at the UHE photon
flux.  Pierre Auger can identify the photons, and it has set an upper limit of
26\% for the fraction of showers caused by primary photo ns at
$10^{19}$eV~\cite{Risse:2005hi}.  This limit will improve significantly in the
near future.  If the diffuse isotropic UHE photon flux is detected, it can
indicate that photomeson interactions of UHECR with CMBR do take place.  A
diffuse anisotropic photon flux with about 10\% increase in the direction of
the galactic center could come from decaying superheavy relic particles in the
galactic halo~\cite{particles}.  However, the spectrum of these photons should
be much harder than that of the photons shown in
Fig.~\ref{fig_spectrum_cutoff}.  This could be used to distinguish between the
two possibilities.  Finally, if UHE photons are detected from a small region
around GC and no diffuse flux is detected, this would mean that photomeson
interactions take place only in the galactic center on starlight photons with
energies $\sim$eV, while no pion production occurs on CMB photons.  This could
be the case, for example, if the Lorentz invariance is broken for high gamma
factors~\cite{lorentz} in such a way that the CMB photons appear below the
threshold for pion production, while starlight photons are above the threshold.

We thank M.~Morris for many helpful discussions.  This work was
supported in part by DOE grant DE-FG03-91ER40662 and NASA grants
ATP02-0000-0151 and ATP03-0000-0057.

\end{document}